\documentclass[fleqn,usenatbib]{mnras}
\usepackage{txfonts}
\usepackage[T1]{fontenc}
\usepackage{ae,aecompl}
\usepackage{graphicx}

\title[Cepheids in Be~51 and Be~55]{New Cepheid variables in the young
open clusters\\ Berkeley~51 and Berkeley~55}

\author[M.~E.~Lohr et al.]{M.~E.~Lohr,$^{1}$\thanks{E-mail: Marcus.Lohr@open.ac.uk}
I.~Negueruela,$^{2}$
H.~M.~Tabernero,$^{2}$
J.~S.~Clark,$^{1}$
F.~Lewis$^{3,4}$
\newauthor
and P.~Roche$^{3}$
\\
$^{1}$School of Physical Sciences, The Open University, Walton Hall,
Milton Keynes MK7~6AA, UK\\
$^{2}$Dpto. de F\'{i}sica, Ingenier\'{i}a de Sistemas y Teor\'{i}a de
la Se\~{n}al, Escuela Polit\'{e}cnica Superior, Universidad de
Alicante,\\
Carretera San Vicente del Raspeig s/n, E-03690 San Vicente
del Raspeig, Spain\\
$^{3}$Faulkes Telescope Project, School of Physics and
Astronomy, Cardiff University, The Parade, Cardiff CF24~3AA, UK\\
$^{4}$Astrophysics Research Institute, Liverpool John Moores University,
146 Brownlow Hill, Liverpool L3~5RF, UK
}

\date{Accepted 2018 May 14. Received 2018 May 11; in original form 2018 April 13}

\pubyear{2018}

\begin{document}
\label{firstpage}
\pagerange{\pageref{firstpage}--\pageref{lastpage}}
\maketitle

\begin{abstract}
As part of a wider investigation of evolved massive stars in Galactic
open clusters, we have spectroscopically identified three candidate
classical Cepheids in the little-studied clusters Berkeley 51,
Berkeley 55 and NGC~6603.  Using new multi-epoch photometry, we
confirm that Be~51 \#162 and Be~55 \#107 are bona fide Cepheids, with
pulsation periods of 9.83$\pm$0.01~d and 5.850$\pm$0.005~d
respectively, while NGC~6603 star W2249 does not show significant
photometric variability.  Using the period-luminosity relationship for
Cepheid variables, we determine a distance to Be~51 of
5.3$^{+1.0}_{-0.8}$~kpc and an age of 44$^{+9}_{-8}$~Myr, placing it
in a sparsely-attested region of the Perseus arm.  For Be~55, we find
a distance of 2.2$\pm$0.3~kpc and age of 63$^{+12}_{-11}$~Myr,
locating the cluster in the Local arm.  Taken together with our recent
discovery of a long-period Cepheid in the starburst cluster VdBH222,
these represent an important increase in the number of young, massive
Cepheids known in Galactic open clusters.  We also consider new Gaia
(data release 2) parallaxes and proper motions for members of Be~51
and Be~55; the uncertainties on the parallaxes do not allow us to
refine our distance estimates to these clusters, but the
well-constrained proper motion measurements furnish further
confirmation of cluster membership.  However, future final Gaia
parallaxes for such objects should provide valuable independent
distance measurements, improving the calibration of the
period-luminosity relationship, with implications for the distance
ladder out to cosmological scales.
\end{abstract}


\section{Introduction}

Evolved stars passing through the instability strip in the
Hertzsprung-Russell diagram can exhibit regular pulsations with
distinctive light curve shapes and periods, allowing their
characterisation as -- amongst others -- $\delta$ Scuti, RR Lyrae or
classical/type I Cepheid variables, according to mass
(e.g. \citealt{chiosi1992}).  The detection of Cepheids in Galactic
open clusters is valuable in several ways: their presence indicates a
relatively young, moderately massive cluster and hence recent star
formation activity in the relevant region of the Galaxy; the brevity
of the yellow supergiant stage makes such objects intrinsically
valuable for constraining models of post-main sequence stellar
evolution; the well-known period-luminosity relationship of Cepheids
\citep{leavitt1912} allows their use as standard candles, providing us
with a distance tracer for the host cluster and hence enhancing our
model of the architecture of the Milky Way; the period-age
relationship for Cepheids \citep{kippenhahn1969} can be independently
checked through isochrone fitting to the whole cluster population; and
finally, future Gaia parallaxes for nearby Cepheids can be used to
produce an improved calibration of the period-luminosity relationship
usable for extragalactic Cepheids, and thus an improved constraint on
the Hubble constant \citep{riess2018}.

Galactic cluster Cepheids are rare: \citet{anderson2013} identified 23
convincing cases in an eight-dimensional all-sky census, and
\citet{chen2015,chen2017} added a further ten, but found that only 31
were usable for constraining the slope of their (near-infrared)
period-luminosity relationship.  Further valid associations between
Cepheids and Galactic open clusters would be of great value.

As part of a search for young open clusters containing evolved stars
in red and yellow super-/hypergiant stages, where extreme mass-loss
rates affect the evolutionary pathways
(e.g. \citealt{clark2009,negueruela2011,dorda2018}), we have
spectroscopically identified a number of candidate Cepheids, and
subsequently undertaken multi-epoch photometry to ascertain their
variability status.  In \citet{clark2015} we confirmed the yellow
supergiant \#505 as a long-period (23.325~d) Cepheid variable in the
starburst cluster VdBH222; here, we report our findings on stars \#162
and \#107, in the faint open clusters Berkeley 51 (Be~51) and Berkeley
55 (Be~55) respectively, in the constellation Cygnus
(\citealt{negueruela2018}, hereafter N18, and
\citealt{negueruela2012}, hereafter N12, respectively).  We also
report on star W2249 in open cluster NGC~6603, in the constellation
Sagittarius.  None of these stars were identified as Cepheids or
variables of any other type in the second Gaia data release.

The three stars are highly probable members of their respective
clusters.  \#162 is in the core of Be~51, as shown in N18 figs.~1 and
2, or 34\arcsec\ from the cluster centre as given on Simbad, where N18
found the cluster to extend to over 3\arcmin\ from the centre.  It
lies on the same isochrone as the spectroscopically-confirmed B-type
cluster members, and the three other F-type supergiants identified in
the cluster core (figs.~10 and 11 in N18).  \#107 also lies right in
the heart of Be~55 (figs.~1 and 2 of N12), at 5\arcsec\ from the
Simbad cluster centre; N12 identified the majority of cluster members
as lying within 3\arcmin\ of the centre, including six of the seven
red or yellow supergiants observed.  Again, it lies on the same
isochrone as B-type confirmed spectroscopic members (figs.~9 and 10 in
N12).  W2249 is 74\arcsec\ from the centre of NGC~6603; the seven
targets -- including W2249 -- for which radial velocities supported
cluster membership \citep{carrera2015} lie within 5\arcmin of the
centre.

\section{Data acquisition and reduction}

\subsection{Spectroscopy}

Star \#162 in Be~51 was observed in the region of the infrared
Ca\,{\sevensize II} triplet on two occasions with the Intermediate
dispersion Spectrograph and Imaging System (ISIS) on the 4.2~m William
Herschel Telescope (WHT), in July 2007 and July 2012, as reported in
N18.  Star \#107 in Be~55 was observed on two occasions with ISIS in
the same spectral region.  The first spectrum, taken in July 2007, is
reported in N12.  The second spectrum was taken with exactly the same
configuration (unbinned \textit{RED}+ CCD, R600R grating and 1\farcs5
slit) on 26 July 2011.  Finally, a higher-resolution spectrum of \#107
in the H$\alpha$ region was taken with the Intermediate Dispersion
Spectrograph (IDS) on the 2.5~m Isaac Newton Telescope (INT) on the
night of 25 September 2017.  The spectrograph was equipped with the
\textit{RED}+2 CCD, the R1200R grating and a 1\farcs5 slit. This
configuration provides a resolving power $R\sim$10\,000 over an
unvignetted range of $\sim$700~\AA, which was centred on $6\,700$~\AA.

Parameters for Be~51 \#162 were derived in N18 using the StePar code
\citep{tabernero2018}; the high-resolution spectrum of Be~55 \#107 was
used to derive basic stellar parameters by employing the same
methodology.  As in that case, we fixed the microturbulence $\xi$
according to the 3D model-based calibration described in
\citet{dutra2016}, while log~$g$ was set to a value of 1.5, typical of
similar objects.

For NGC~6603, archival spectra from \citet{carrera2015} for radial
velocity likely cluster members were downloaded and re-reduced to
search for evidence of candidate Cepheids.

\subsection{Photometry}

Time series photometry was obtained for the three clusters using the
Las Cumbres Observatory (LCO), described in
\citet{brown2013}\footnote{Recent changes to LCO instruments and data
  products can be found at https://lco.global/observatory/}.  For
Be~51, 41 usable observations were made between 23 May and 19
September 2015, with both Bessell $V$ and SDSS $i'$ filters (30~s
exposures, pixel scale 0\farcs301~pixel\textsuperscript{-1}).  For
Be~55, 14 initial observations were made in $R$ between 30 June and 29
July 2017 (10~s exposures, 0\farcs301~pixel\textsuperscript{-1});
follow-up observations occurred between 7 October and 5 November: 18
epochs in $R$ and 15 in $V$ (30~s exposures,
0\farcs387~pixel\textsuperscript{-1}). For NGC~6603, 20 usable
observations were made in $R$ between 11 and 30 July 2017 (10~s
exposures, 0\farcs304~pixel\textsuperscript{-1}).

Basic reductions including bad pixel masking, bias and dark
subtraction, and flat field correction, were performed using the LCOGT
data pipeline.  For each cluster, sets of images with a shared pixel
scale were rotated if necessary, realigned and trimmed to a common
coordinate system and area using the IRAF tasks $geomap$ and
$imalign$.  Point spread function (PSF) fitting photometry was then
carried out using the IRAF/DAOPHOT package.  In each group of images,
one frame judged to be of excellent quality -- small measured full
width at half maximum (FWHM) of the PSF of isolated bright targets,
good signal to noise ratio, absence of artefacts -- was initially
processed to determine the locations of genuine point sources; these
coordinates were then used as the starting point for processing all
other frames.

After tests with various numbers of PSF stars, functional forms of the
analytic component of the PSF model, and orders of empirical
variability, the best results were achieved using a three-parameter
elliptical Moffat function with $\beta$ = 2.5, and an empirical
constant PSF model; five PSF stars were selected for each cluster as
close to the targets of interest as possible, and covering a
comparable range of brightness.  Since the observations for each
cluster and filter had been made over many nights, under different
conditions and often with different instruments, it was necessary to
determine the characteristics of each frame individually (FWHM of the
PSF of isolated bright stars, sky level, and standard deviation of the
sky level) in order to achieve acceptable PSF modelling.

The mid-times of observation of each frame were then converted to
BJD(TDB)\footnote{http://astroutils/astronomy.ohio-state.edu/time/
  (see also \citealp{eastman2010})}.  Using these, light curves could
be constructed for all targets believed to be cluster members and
bright enough to have magnitudes measured in every frame.  Plotting
all these together revealed both the typical night-to-night variations
associated with changing observing conditions, and the presence of
intrinsically variable stars.  The light curves of a subset of
non-variable objects believed to possess similar spectral types to the
candidate Cepheid in each cluster were then combined to produce a
reference star, relative to which a differential light curve could be
constructed for each suspected variable object.  This method resulted
in smoother final light curves, with lower uncertainties, than using a
single reference star.

\section{Results and Discussion}

\subsection{Be~51}

\begin{figure}
\includegraphics[width=\columnwidth]{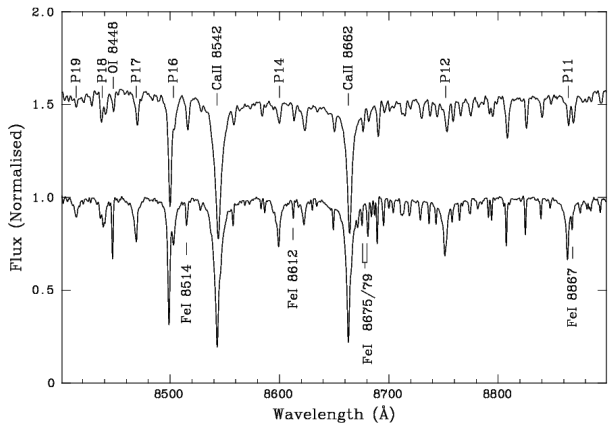}
\caption{ISIS spectra at two epochs for Be~51 \#162 showing changes in
  spectral type.  The upper spectrum is from July 2007, corresponding
  to an early-G type; the lower spectrum, from July 2012 (note the
  higher resolution), is around F8\,Ib.  The Paschen and
  O\,{\sevensize I}~8446~\AA\ lines weaken significantly as we move to
  G types.  Conversely, the general metallic spectrum (some of the
  strongest Fe\,{\sevensize I} lines are marked) becomes stronger.
  The apparent inversion of the ratio between the Fe\,{\sevensize I}
  lines at 8675 and 8679~\AA\ is due to the disappearance of
  N\,{\sevensize I}~8680~\AA, the strongest of a group of
  N\,{\sevensize I} lines that characterise the spectra of A and
  F-type stars in this spectral region.}
\label{Be51specs}
\end{figure}

Spectral variations between the two epochs for suspected Cepheid \#162
are evident, despite the different resolution, as illustrated in
Fig.~\ref{Be51specs}.  Parameters for \#162 were derived in N18, where
it was found to have a very slightly supersolar metallicity.

\begin{figure}
\includegraphics[width=\columnwidth]{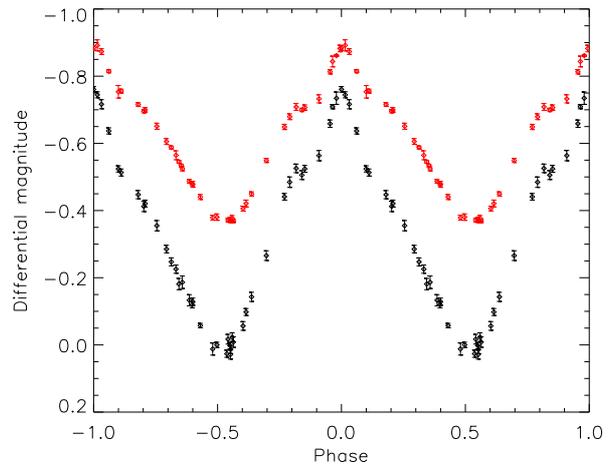}
\caption{Differential light curves for \#162 in $V$ (black, lower
  curve) and $i'$ (red, upper curve), folded on $P$ = 9.83~d, with phase
  zero set from the $V$-band maximum.  An artificial offset of 0.2 mag
  between the two curves has been inserted for clarity.}
\label{Be51}
\end{figure}

Initial inspection of the time series photometry for Be~51 revealed
clear variability in the light curve of \#162, relative to the light
curves of other cluster members.  No other cluster members studied
showed obvious intrinsic variability.  A differential light curve was
constructed for \#162 relative to the combined light curve of four
other cluster supergiants (\#105, \#134, \#146 and \#172), with mid-F
or early K classifications in N18.  A period of 9.83$\pm$0.01~d was
determined for \#162 by a form of string length minimization
(e.g. \citealp{dworetsky1983}), a method well-suited to small
quantities of data where the shape of the light curve may not be
sinusoidal; however, checks using Lomb-Scargle periodograms
\citep{lomb1976,scargle1982,horne1986} and phase dispersion
minimization (e.g. \citealp{lafler1965,stellingwerf1978}) confirmed
this as the best period in the range 0.5--100~d, and as exceeding the
false alarm probability (FAP) 1\% threshold.  Fig.~\ref{Be51} shows
\#162's light curves in $V$ and $i'$ folded on this best period.
These have the expected shape (according to the Hertzsprung
progression, \citealp{hertzsprung1926}) of a type I Cepheid of period
$\sim$10~d, with bumps on both ascending and descending branches
(compare also fig.~1 in \citealp{soszynski2008}).  Maxima were
obtained at BJD 2\,457\,202.8815 in $V$ and 2\,457\,202.8823 in $i'$.

The $V$-band photometry for \#162 was calibrated using the
standardized photometric results from N18, allowing us to determine an
average observed magnitude for the Cepheid of $\langle m_V \rangle$ =
15.295$\pm$0.008 (range was 14.902$\pm$0.006 to 15.688$\pm$0.008).  An
average absolute $V$-band magnitude was derived from the pulsation
period using Eqn.~19 in \citet{anderson2013}: $\langle M_V \rangle$ =
$-$3.88$\pm$0.24.  An $E$($B-V$) = 1.79$\pm$0.09 for \#162 was
estimated as the mean of the reddenings calculated in N18 for eleven
spectroscopically-confirmed B-type cluster members (for comparison,
$\langle E$($B-V$)$ \rangle$ = 1.76$\pm$0.12 was estimated for seven
supergiants with photometry); with $R_V$ = 3.1 we then obtain $A_V$ =
5.55$\pm$0.28.  Thus we may calculate a distance to the cluster of
5.3$^{+1.0}_{-0.8}$~kpc, where the uncertainties are dominated by the
uncertainty in the extinction.  Using an alternative calibration of
the period-luminosity relationship based on Hubble parallaxes for
Galactic Cepheids \citep{benedict2007}, the distance is
5.7$^{+0.8}_{-0.7}$~kpc.  Employing the period-age relationship for
fundamental-mode Cepheids in \citet{bono2005}, an age of
44$^{+9}_{-8}$~Myr is obtained.

These values are consistent with the preferred distance of
$\sim$5.5~kpc found for Be~51 by N18 on the basis of cluster
photometry and radial velocities, and with their preferred age for
this distance of $\sim$60~Myr using isochrone fitting to a dereddened
colour-magnitude diagram for probable cluster members.  In contrast to
earlier estimates based on photometry alone, which regarded Be~51 as a
much older, closer cluster within the Local arm
\citep{tadross2008,subramaniam2010,kharchenko2013}, a distance of
5.3~kpc with $\ell$ = 72\fdg147 would seem to place it in the Perseus
arm, in a region lacking in reliable distance tracers (see
e.g. fig.~11 in \citealp{zhang2013}, fig.~14 in \citealp{choi2014} and
fig.~5 in \citealp{reid2016}).

\subsection{Be~55}

\begin{figure}
\includegraphics[width=\columnwidth]{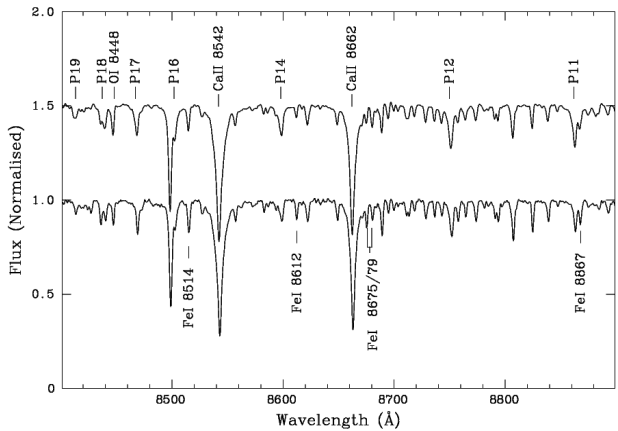}
\caption{ISIS spectra at two epochs for Be~55 \#107 showing changes in
  spectral type.  The upper spectrum, from July 2007, presents a
  spectral type F8\,Ib, while the lower spectrum, taken in July 2011,
  is early G.  Changes are similar to those seen in Be~51 \#162
  (Fig.~\ref{Be51specs}).}
\label{Be55specs}
\end{figure}

Again, for Cepheid candidate \#107, significant changes in spectral
type are seen between the two epochs (see Fig.~\ref{Be55specs}).  We
also find, for the high-resolution spectrum, $T_{{\rm eff}}$ =
5\,505$\pm$199~K and [M/H] = 0.07$\pm$0.12, fully consistent with a
solar metallicity.

The raw light curves for Be~55 supported significant variability in
\#107, but also in \#198, classified in N12 as a Be shell star (see
Fig.~8 in N12).  Construction of a differential light curve for \#198
relative to various subsets of other bright cluster members did not,
however, reveal any significant periodicity to this variation, so it
may be produced by some aspect of the Be phenomenon
\citep{rivinius2013} rather than, for example, an eclipsing binary.

\begin{figure}
\includegraphics[width=\columnwidth]{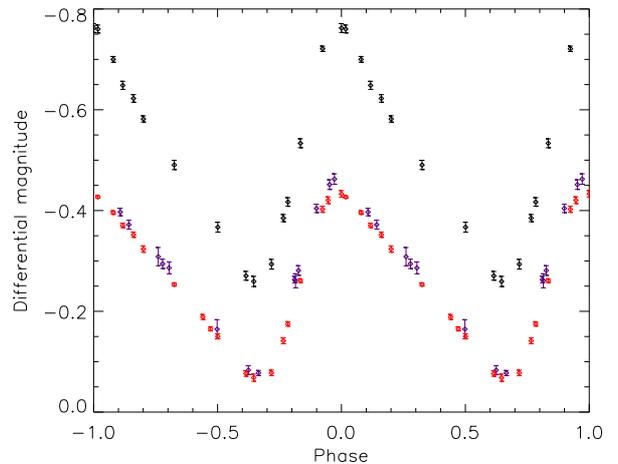}
\caption{Differential light curves for \#107 in $V$ (black, upper
  curve) and $R$ (purple and red, lower curves), folded on $P$ = 5.85~d.
  The small vertical offset between the two $R$-band curves and their
  different uncertainty sizes are caused by the different exposure
  lengths and pixel scales of the two sets of observations.}
\label{Be55}
\end{figure}

The differential light curve for \#107 was constructed relative to the
combined light curve of four K and G supergiants (\#110, \#145, \#163
and \#196 from N12), and its period was determined as
5.850$\pm$0.005~d by string length minimization again.  This was also
confirmed as the most significant periodicity over the range
0.5--100~d by Lomb-Scargle and phase dispersion minimization methods,
and far surpassed the 1\% FAP threshold.  Fig.~\ref{Be55} shows the
$V$ and $R$-band light curves folded on this period; again, it is
apparent that they have the shape of a type I Cepheid with pulsation
period $\sim$5~d, without bumps on either branch, and with a linear
descending branch.  Maxima were observed at BJD 2\,458\,053.6674 in
$V$ and 2\,458\,053.6664 in $R$.

The $V$-band light curve was calibrated using the photometry of N12,
giving $\langle m_V \rangle$ = 13.834$\pm$0.008 (range was
13.583$\pm$0.006 to 14.085$\pm$0.009).  Using the same approach as for
Be~51, $\langle M_V \rangle$ was determined as $-$3.23$\pm$0.21, and
$E$($B-V$) as 1.74$\pm$0.07 (the mean of the reddenings calculated in
N12 for seven spectroscopically-confirmed B-type cluster members
excluding the Be shell star \#198); this gave $A_V$ = 5.39$\pm$0.22.
Our calculated distance to the cluster is therefore 2.2$\pm$0.3~kpc,
or using the relationship of \citet{benedict2007},
2.4$^{+0.3}_{-0.2}$~kpc, with an age of 63$^{+12}_{-11}$~Myr.

This distance is somewhat less than the 4.0$\pm$0.6~kpc obtained by
N12 by a visual fit to the zero-age main sequence on a dereddened
$M_V$/$B-V_0$ diagram for probable cluster members.  However, their
age estimate of $\sim$50~Myr is compatible with ours.  Moreover, given
the evidence supporting \#107's membership of Be~55 given in N12,
including the central location of \#107 within the cluster, and the
presence of five other late-type supergiants in close proximity, we
feel it is more likely that this apparent mismatch is caused by
underestimated uncertainties in N12's distance modulus (determined by
a single method, rather than the multiple independent approaches
discussed in N18), than by an unrelated Cepheid coincidentally lying
along our line of sight to the cluster.  Earlier purely photometric
studies \citep{maciejewski2007,tadross2008} had found a even lower
distance (1.2~kpc) and much greater age ($\sim$300~Myr), which N12
notes is incompatible with the observed population of B3--4 stars.
Our distance of 2.2~kpc with $\ell$ = 93\fdg027 would seem to locate
Be~55 on the outer edge of the Local arm (\citealp{xu2013}, especially
fig.~10), rather than in the Perseus arm as N12 suggest.

\subsection{NGC~6603}

The re-reduced spectrum of object W2249 showed an early-G spectral
type, similar to the candidate Cepheids in Be~51 and Be~55, motivating
further photometric observations.  However, no significant variability
was detected in its differential light curve, constructed relative to
the combined light curves of six other candidate members of NGC~6603
with similar $V$ magnitudes (W1997, W2033, W2215, W2252, W2352 and
W2438).  The full amplitude of variability exhibited was
$\sim$0.02~mag in $R$, comparable to the size of the photometric
uncertainties.

We may note that age estimates for this cluster are highly
inconsistent, ranging from $\sim$60~Myr \citep{kharchenko2005} to
$\sim$500~Myr \citep{sagar1998}; ages above $\sim$200~Myr would place
it outside a plausible mass range for Cepheids.  (This uncertainty in
age may be explained by an inappropriate assumption of solar
metallicity for the cluster; \citet{carrera2015} found NGC~6603 to be
one of the most metal-rich open clusters known.)

\subsection{Gaia data release 2 (DR2)}

\begin{figure}
\includegraphics[width=\columnwidth]{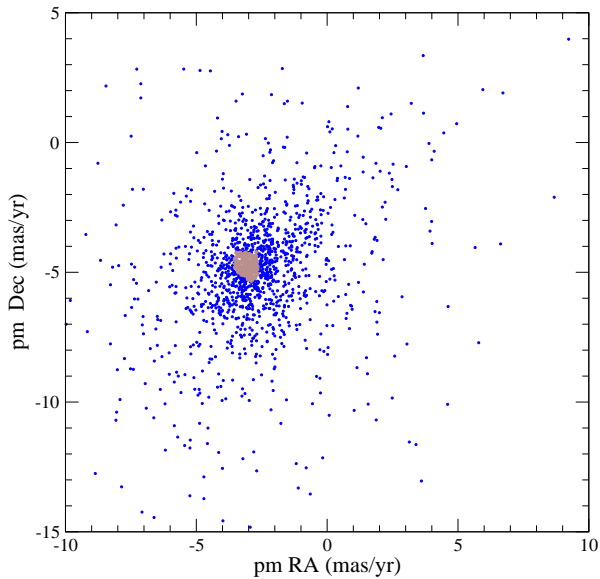}
\caption{Gaia DR2 proper motions for targets within 3\arcmin of the
  centre of Be~51, showing concentration of selected candidate members
  in grey.}
\label{Be51sel}
\end{figure}

\begin{figure}
\includegraphics[width=\columnwidth]{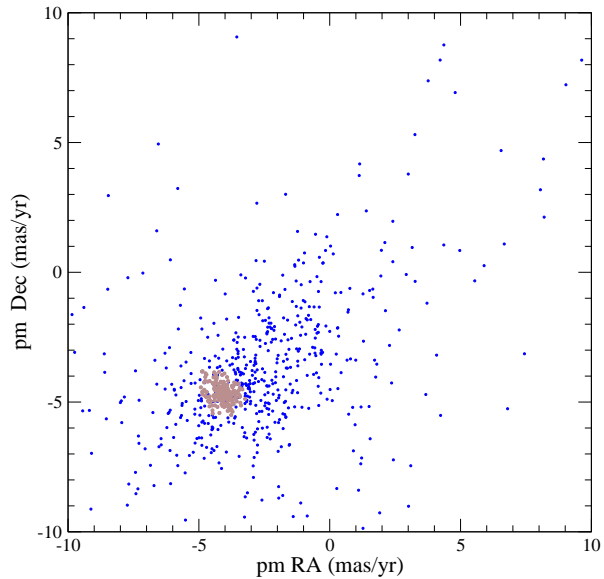}
\caption{Gaia DR2 proper motions for targets within 3\farcm5 of the
  centre of Be~55, showing concentration of selected candidate members
  in grey.  This field has a much lower stellar density than that of
  Be~51, but the cluster members are about two magnitudes brighter,
  resulting in smaller uncertainties.}
\label{Be55sel}
\end{figure}

\begin{figure}
\includegraphics[width=\columnwidth]{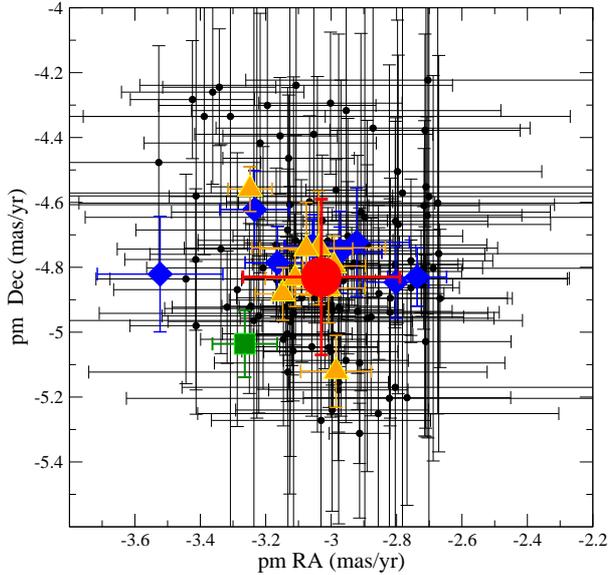}
\caption{Gaia DR2 proper motions, with uncertainties, for cleaned
  sample of 117 stars identified as probable members of Be~51.  The
  Cepheid \#162 is shown with a green square; the eight other cool
  supergiants (listed in the top panel of N18 table 3) are shown with
  yellow triangles; B-type stars observed spectroscopically are shown
  with blue diamonds.  One B-type star, \#143, is not included here or
  in Fig.~\ref{Be51px} because its astrometric solution in DR2 appears
  faulty (it has a negative parallax with very large associated
  uncertainty: $-$0.36$\pm$0.21).  The large red circle indicates the
  weighted average for the cleaned sample, with error bars
  corresponding to the median uncertainty for single stars: pmRA =
  $-$3.03$\pm$0.20~mas~y\textsuperscript{-1}, pmDec =
  $-$4.83$\pm$0.24~mas~y\textsuperscript{-1}.}
\label{Be51pm}
\end{figure}

\begin{figure}
\includegraphics[width=\columnwidth]{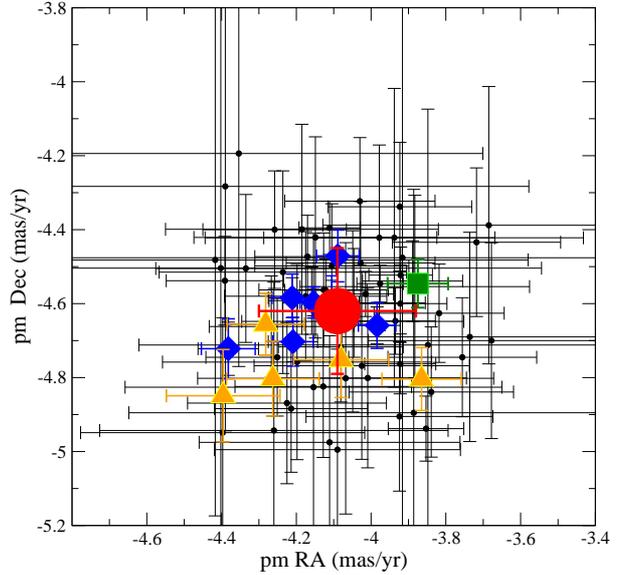}
\caption{Gaia DR2 proper motions, with uncertainties, for cleaned
  sample of 44 stars identified as probable members of Be~55.  The
  Cepheid \#107 is shown with a green square; the five other cool
  supergiants (listed in N12 table 7, excluding possible foreground
  interloper S61) are shown with yellow triangles; B-type stars
  observed spectroscopically are shown with blue diamonds.  One B-type
  star, \#129, is not found in DR2.  The large red circle indicates
  the weighted average for the cleaned sample, with error bars
  corresponding to the median uncertainty for single stars: pmRA =
  $-$4.09$\pm$0.19~mas~y\textsuperscript{-1}, pmDec =
  $-$4.62$\pm$0.18~mas~y\textsuperscript{-1}.}
\label{Be55pm}
\end{figure}

The second Gaia data release \citep{gaia2018} made available precise
positions, parallaxes and proper motions for most of the stars
previously identified as probable members of Be~51 and Be~55.
Therefore, we used DR2 data to investigate these clusters afresh.  As
shown in Figs.~\ref{Be51sel} and \ref{Be55sel}, the two clusters
appear as clear overdensities in the proper motion (pmRA/pmDec) plane,
allowing an initial selection of possible cluster members.  We then
calculated the average proper motion for each cluster, weighting
values with the inverse of their uncertainties.  Each sample was
cleaned iteratively, by discarding outliers and recalculating the
average, until the standard deviation of the sample's proper motion
was comparable with the median error on an individual value.  (Removal
of outliers does not imply any judgement on their cluster membership,
but simply allows us to define a clean sample of objects with
moderately low errors.  The procedure is very robust, as the weighted
averages do not change significantly throughout.)  Figs.~\ref{Be51pm}
and \ref{Be55pm} show the results for these cleaned samples, with the
values for B-type and supergiant stars identified spectroscopically in
N12 and N18 highlighted.  These reveal that the proper motions fall
within a very narrow range for each cluster, and further support
cluster membership for almost all of the spectroscopic targets,
including, notably, the two Cepheids Be~51 \#162 and Be~55 \#107.

However, when we investigate parallaxes for these cluster stars, a
number of limitations and warnings regarding the DR2 astrometry must
be borne in mind, as outlined in
\citet{lindegren2018,arenou2018,luri2018}.  A parallax zero point of
$\sim$$-$0.03~mas is found from observations of over half a million
quasars (and indeed, comparison with a sample of eclipsing binaries
with accurate distances suggests a larger zero point of $-$0.08~mas
\citep{stassun2018}, although the use of a single-star model for all
DR2 stars may increase the difference for binary systems).  Moreover,
there are spatial correlations in parallax and proper motion on scales
$\sim$1\degr, a number of negative and spurious parallaxes, and
parallax uncertainties underestimated by up to 50\% for sources in the
$G$-band magnitude range 12-15.  As a consequence, individual
parallaxes for stars beyond 1~kpc are unsafe, and averaging parallaxes
over the whole population of an open cluster will not reduce the
uncertainty on the mean beyond the $\pm$0.1~mas level.

\begin{figure}
\includegraphics[width=\columnwidth]{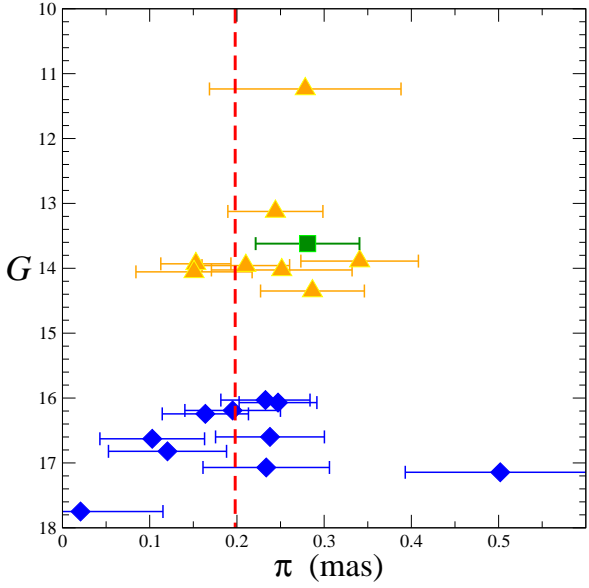}
\caption{Gaia DR2 parallaxes, with uncertainties, against $G$-band
  mean magnitude for spectroscopically-identified B-type and
  supergiant stars in Be~51 (colours and labels as in
  Fig.~\ref{Be51pm}).  The red dashed line indicates the weighted
  average parallax for the whole cleaned sample: $\pi$ = 0.20~mas,
  with a standard deviation for the sample of 0.09~mas.  The B-type
  star with the largest parallax is \#153, which is also on the edge
  of the proper motion distribution in Fig.~\ref{Be51pm}; this object
  could be a non-cluster member, though its astrometric solution may
  have suffered from the presence of a close companion.}
\label{Be51px}
\end{figure}

\begin{figure}
\includegraphics[width=\columnwidth]{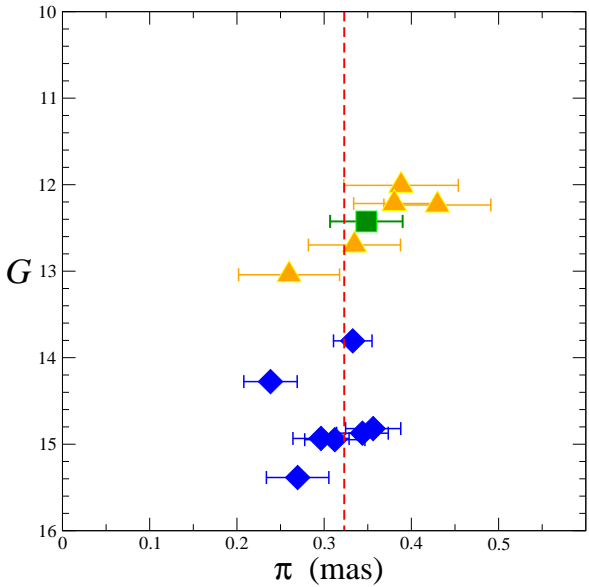}
\caption{Gaia DR2 parallaxes, with uncertainties, against $G$-band
  mean magnitude for spectroscopically-identified B-type and
  supergiant stars in Be~55 (colours and labels as in
  Fig.~\ref{Be55pm}).  The red dashed line indicates the weighted
  average parallax for the whole cleaned sample: $\pi$ = 0.32~mas,
  with a standard deviation for the sample of 0.08~mas.}
\label{Be55px}
\end{figure}

With these caveats in mind, we calculated the average parallax for
each cleaned cluster sample, again weighting each measurement with the
inverse of its uncertainty, and removed stars which were incompatible
with these average values within their respective uncertainties.  (As
before, removal of a star does not imply that it is not a cluster
member, although many of the removed objects may be expected to be
either background or foreground objects unless their errors are very
strongly underestimated.  Again, the weighted averages are not
significantly changed by the cleaning process.)  Figs.~\ref{Be51px}
and \ref{Be55px} show the results, plotting only the
previously-identified supergiants and B-type stars along with the
(full) sample average parallaxes.  It is notable that the parallaxes
of these cluster members are much more widely scattered than their
proper motions, and that the supergiants tend to have larger
parallaxes than the blue stars, which are concentrated around the
cluster averages.  Since all objects should be compatible with the
average, the uncertainties are clearly underestimated.

\citet{luri2018} advise against inverting DR2 parallaxes, especially
for individual objects, to obtain distances, and instead recommend the
use of Bayesian inference for this purpose.  However, given the
substantial scatter in the parallaxes of Be~51 and Be~55 members, the
acknowledged significant underestimation of parallax uncertainties for
stars in their magnitude range, and the high and variable extinction
in these clusters, we feel such an approach would be of limited value
at this stage.  So, taken at face value (i.e. using simple inversion
of the cleaned cluster sample average parallax), the distance to Be~51
would be $\sim$5~kpc, with the uncertainties implying a range between
3.3 and 10~kpc.  For Be~55, the nominal distance is 3.1~kpc, with an
implied range of 2.4-4.5~kpc.  Hence, although these distances are
consistent with those found earlier in this work and in N12 and N18,
we cannot at this stage use Gaia DR2 parallaxes to determine distances
to Be~51 and Be~55 which are more reliable or precise than those found
by other methods.

\section{Conclusions}

From spectroscopic and photometric variability, we have confirmed that
Be~51 \#162 and Be~55 \#107 are bona fide classical Cepheids.  For
\#162, we determine a pulsation period of 9.83$\pm$0.01~d, implying a
distance to Be~51 of 5.3$^{+1.0}_{-0.8}$~kpc (5.7$^{+0.8}_{-0.7}$~kpc
using another period-luminosity calibration) and an age of
44$^{+9}_{-8}$~Myr, consistent with values found independently in N18,
and placing the cluster in a sparsely-known region of the Perseus arm.
For \#107, we find $P$ = 5.850$\pm$0.005~d, and hence for Be~55, $d$ =
2.2$\pm$0.3~kpc (2.4$^{+0.3}_{-0.2}$~kpc) and age =
63$^{+12}_{-11}$~Myr.  This distance would place the cluster in the
Local arm.  The ages determined for both clusters are also in the
interesting range $\sim$50--60~Myr; as noted in N12, it is rare to
find red supergiants in older clusters, so the K supergiants
identified in Be~51 and Be~55 may provide valuable data on the
boundary between intermediate and massive stars.

Taken together with our Cepheid discovery in \citet{clark2015},
locating the starburst cluster VdBH222 unexpectedly in or near the
inner 3~kpc Galactic arm, these new Cepheids in young/intermediate age
clusters provide a richer understanding of the architecture of the
Milky Way and its recent star formation history.  They also represent
an important increase in the number of young, massive Cepheids known
in Galactic open clusters.  While the recent Gaia DR2 parallaxes for
members of these clusters do not yet allow a reliable check on the
distances determined here, future Gaia results should provide an
independent determination of the distances to such Cepheids and their
host clusters, and thereby improve the calibration of the
period-luminosity relationship, with implications for the distance
ladder out to cosmological scales.

\section*{Acknowledgements}

The Faulkes Telescope Project is an education partner of LCO.  The
Faulkes Telescopes are maintained and operated by LCO.  The WHT and
INT are operated on the island of La Palma by the Isaac Newton Group
in the Spanish Observatorio del Roque de los Muchachos of the
Instituto de Astrof\'{i}sica de Canarias.  This work has made use of
the WEBDA database, operated at the Department of Theoretical Physics
and Astrophysics of the Masaryk University.  This research is
partially supported by the Spanish Government Ministerio de
Econom\'{i}a y Competitivad (MINECO/FEDER) under grants
FJCI-2014-23001 and AYA2015-68012-C2-2-P, and by the UK Science and
Technology Facilities Council under grant ST/P000584/1.  We thank Dr
Carlos Gonz\'{a}lez Fern\'{a}ndez for obtaining and reducing the 2011
spectrum for Be~55 \#107.

\bibliographystyle{mnras}
\bibliography{cephrefsv3}

\begin{thebibliography}{}
\makeatletter
\relax
\def\mn@urlcharsother{\let\do\@makeother \do\$\do\&\do\#\do\^\do\_\do\%\do\~}
\def\mn@doi{\begingroup\mn@urlcharsother \@ifnextchar [ {\mn@doi@}
  {\mn@doi@[]}}
\def\mn@doi@[#1]#2{\def\@tempa{#1}\ifx\@tempa\@empty \href
  {http://dx.doi.org/#2} {doi:#2}\else \href {http://dx.doi.org/#2} {#1}\fi
  \endgroup}
\def\mn@eprint#1#2{\mn@eprint@#1:#2::\@nil}
\def\mn@eprint@arXiv#1{\href {http://arxiv.org/abs/#1} {{\tt arXiv:#1}}}
\def\mn@eprint@dblp#1{\href {http://dblp.uni-trier.de/rec/bibtex/#1.xml}
  {dblp:#1}}
\def\mn@eprint@#1:#2:#3:#4\@nil{\def\@tempa {#1}\def\@tempb {#2}\def\@tempc
  {#3}\ifx \@tempc \@empty \let \@tempc \@tempb \let \@tempb \@tempa \fi \ifx
  \@tempb \@empty \def\@tempb {arXiv}\fi \@ifundefined
  {mn@eprint@\@tempb}{\@tempb:\@tempc}{\expandafter \expandafter \csname
  mn@eprint@\@tempb\endcsname \expandafter{\@tempc}}}

\bibitem[\protect\citeauthoryear{{Anderson}, {Eyer}  \& {Mowlavi}}{{Anderson}
  et~al.}{2013}]{anderson2013}
{Anderson} R.~I.,  {Eyer} L.,   {Mowlavi} N.,  2013, \mn@doi [\mnras]
  {10.1093/mnras/stt1160}, 434, 2238

\bibitem[\protect\citeauthoryear{{Arenou} et~al.,}{{Arenou}
  et~al.}{2018}]{arenou2018}
{Arenou} F.,  et~al., 2018, preprint (\mn@eprint {arXiv} {1804.09375})

\bibitem[\protect\citeauthoryear{{Benedict} et~al.,}{{Benedict}
  et~al.}{2007}]{benedict2007}
{Benedict} G.~F.,  et~al., 2007, \mn@doi [\aj] {10.1086/511980}, 133, 1810

\bibitem[\protect\citeauthoryear{{Bono}, {Marconi}, {Cassisi}, {Caputo},
  {Gieren}  \& {Pietrzynski}}{{Bono} et~al.}{2005}]{bono2005}
{Bono} G.,  {Marconi} M.,  {Cassisi} S.,  {Caputo} F.,  {Gieren} W.,
  {Pietrzynski} G.,  2005, \mn@doi [\apj] {10.1086/427744}, 621, 966

\bibitem[\protect\citeauthoryear{{Brown} et~al.,}{{Brown}
  et~al.}{2013}]{brown2013}
{Brown} T.~M.,  et~al., 2013, \mn@doi [\pasp] {10.1086/673168}, 125, 1031

\bibitem[\protect\citeauthoryear{{Carrera}, {Casamiquela}, {Ospina},
  {Balaguer-N{\'u}{\~n}ez}, {Jordi}  \& {Monteagudo}}{{Carrera}
  et~al.}{2015}]{carrera2015}
{Carrera} R.,  {Casamiquela} L.,  {Ospina} N.,  {Balaguer-N{\'u}{\~n}ez} L.,
  {Jordi} C.,   {Monteagudo} L.,  2015, \mn@doi [\aap]
  {10.1051/0004-6361/201425531}, 578, A27

\bibitem[\protect\citeauthoryear{{Chen}, {de Grijs}  \& {Deng}}{{Chen}
  et~al.}{2015}]{chen2015}
{Chen} X.,  {de Grijs} R.,   {Deng} L.,  2015, \mn@doi [\mnras]
  {10.1093/mnras/stu2165}, 446, 1268

\bibitem[\protect\citeauthoryear{{Chen}, {de Grijs}  \& {Deng}}{{Chen}
  et~al.}{2017}]{chen2017}
{Chen} X.,  {de Grijs} R.,   {Deng} L.,  2017, \mn@doi [\mnras]
  {10.1093/mnras/stw2390}, 464, 1119

\bibitem[\protect\citeauthoryear{{Chiosi}, {Bertelli}  \& {Bressan}}{{Chiosi}
  et~al.}{1992}]{chiosi1992}
{Chiosi} C.,  {Bertelli} G.,   {Bressan} A.,  1992, \mn@doi [\araa]
  {10.1146/annurev.aa.30.090192.001315}, 30, 235

\bibitem[\protect\citeauthoryear{{Choi}, {Hachisuka}, {Reid}, {Xu},
  {Brunthaler}, {Menten}  \& {Dame}}{{Choi} et~al.}{2014}]{choi2014}
{Choi} Y.~K.,  {Hachisuka} K.,  {Reid} M.~J.,  {Xu} Y.,  {Brunthaler} A.,
  {Menten} K.~M.,   {Dame} T.~M.,  2014, \mn@doi [\apj]
  {10.1088/0004-637X/790/2/99}, 790, 99

\bibitem[\protect\citeauthoryear{{Clark} et~al.,}{{Clark}
  et~al.}{2009}]{clark2009}
{Clark} J.~S.,  et~al., 2009, \mn@doi [\aap] {10.1051/0004-6361/200911945},
  498, 109

\bibitem[\protect\citeauthoryear{{Clark}, {Negueruela}, {Lohr}, {Dorda},
  {Gonz{\'a}lez-Fern{\'a}ndez}, {Lewis}  \& {Roche}}{{Clark}
  et~al.}{2015}]{clark2015}
{Clark} J.~S.,  {Negueruela} I.,  {Lohr} M.~E.,  {Dorda} R.,
  {Gonz{\'a}lez-Fern{\'a}ndez} C.,  {Lewis} F.,   {Roche} P.,  2015, \mn@doi
  [\aap] {10.1051/0004-6361/201527360}, 584, L12

\bibitem[\protect\citeauthoryear{{Dorda}, {Negueruela}  \&
  {Gonz\'{a}lez-Fern\'{a}ndez}}{{Dorda} et~al.}{2018}]{dorda2018}
{Dorda} R.,  {Negueruela} I.,   {Gonz\'{a}lez-Fern\'{a}ndez} C.,  2018, \mn@doi
  [\mnras] {10.1093/mnras/stx3317}, 475, 2003

\bibitem[\protect\citeauthoryear{{Dutra-Ferreira}, {Pasquini}, {Smiljanic},
  {Porto de Mello}  \& {Steffen}}{{Dutra-Ferreira} et~al.}{2016}]{dutra2016}
{Dutra-Ferreira} L.,  {Pasquini} L.,  {Smiljanic} R.,  {Porto de Mello} G.~F.,
   {Steffen} M.,  2016, \mn@doi [\aap] {10.1051/0004-6361/201526783}, 585, A75

\bibitem[\protect\citeauthoryear{{Dworetsky}}{{Dworetsky}}{1983}]{dworetsky198%
3}
{Dworetsky} M.~M.,  1983, \mn@doi [\mnras] {10.1093/mnras/203.4.917}, 203, 917

\bibitem[\protect\citeauthoryear{{Eastman}, {Siverd}  \& {Gaudi}}{{Eastman}
  et~al.}{2010}]{eastman2010}
{Eastman} J.,  {Siverd} R.,   {Gaudi} B.~S.,  2010, \mn@doi [\pasp]
  {10.1086/655938}, 122, 935

\bibitem[\protect\citeauthoryear{{Gaia Collaboration}, {Brown}, {Vallenari},
  {Prusti}, {de Bruijne}, {Babusiaux}  \& {Bailer-Jones}}{{Gaia Collaboration}
  et~al.}{2018}]{gaia2018}
{Gaia Collaboration}, {Brown} A.~G.~A.,  {Vallenari} A.,  {Prusti} T.,  {de
  Bruijne} J.~H.~J.,  {Babusiaux} C.,   {Bailer-Jones} C.~A.~L.,  2018,
  preprint (\mn@eprint {arXiv} {1804.09365})

\bibitem[\protect\citeauthoryear{{Hertzsprung}}{{Hertzsprung}}{1926}]{hertzspr%
ung1926}
{Hertzsprung} E.,  1926, \bain, 3, 115

\bibitem[\protect\citeauthoryear{{Horne} \& {Baliunas}}{{Horne} \&
  {Baliunas}}{1986}]{horne1986}
{Horne} J.~H.,  {Baliunas} S.~L.,  1986, \mn@doi [\apj] {10.1086/164037}, 302,
  757

\bibitem[\protect\citeauthoryear{{Kharchenko}, {Piskunov}, {R\"{o}ser},
  {Schilbach}  \& {Scholz}}{{Kharchenko} et~al.}{2005}]{kharchenko2005}
{Kharchenko} N.~V.,  {Piskunov} A.~E.,  {R\"{o}ser} S.,  {Schilbach} E.,
  {Scholz} R.-D.,  2005, \mn@doi [\aap] {10.1051/0004-6361:20042523}, 438, 1163

\bibitem[\protect\citeauthoryear{{Kharchenko}, {Piskunov}, {Schilbach},
  {R{\"o}ser}  \& {Scholz}}{{Kharchenko} et~al.}{2013}]{kharchenko2013}
{Kharchenko} N.~V.,  {Piskunov} A.~E.,  {Schilbach} E.,  {R{\"o}ser} S.,
  {Scholz} R.-D.,  2013, \mn@doi [\aap] {10.1051/0004-6361/201322302}, 558, A53

\bibitem[\protect\citeauthoryear{{Kippenhahn} \& {Smith}}{{Kippenhahn} \&
  {Smith}}{1969}]{kippenhahn1969}
{Kippenhahn} R.,  {Smith} L.,  1969, \aap, 1, 142

\bibitem[\protect\citeauthoryear{{Lafler} \& {Kinman}}{{Lafler} \&
  {Kinman}}{1965}]{lafler1965}
{Lafler} J.,  {Kinman} T.~D.,  1965, \mn@doi [\apjs] {10.1086/190116}, 11, 216

\bibitem[\protect\citeauthoryear{{Leavitt} \& {Pickering}}{{Leavitt} \&
  {Pickering}}{1912}]{leavitt1912}
{Leavitt} H.~S.,  {Pickering} E.~C.,  1912, Harvard College Observatory
  Circular, 173, 1

\bibitem[\protect\citeauthoryear{{Lindegren} et~al.,}{{Lindegren}
  et~al.}{2018}]{lindegren2018}
{Lindegren} L.,  et~al., 2018, preprint (\mn@eprint {arXiv} {1804.09366})

\bibitem[\protect\citeauthoryear{{Lomb}}{{Lomb}}{1976}]{lomb1976}
{Lomb} N.~R.,  1976, \mn@doi [\apss] {10.1007/BF00648343}, 39, 447

\bibitem[\protect\citeauthoryear{{Luri} et~al.,}{{Luri}
  et~al.}{2018}]{luri2018}
{Luri} X.,  et~al., 2018, preprint (\mn@eprint {arXiv} {1804.09376})

\bibitem[\protect\citeauthoryear{{Maciejewski} \& {Niedzielski}}{{Maciejewski}
  \& {Niedzielski}}{2007}]{maciejewski2007}
{Maciejewski} G.,  {Niedzielski} A.,  2007, \mn@doi [\aap]
  {10.1051/0004-6361:20066588}, 467, 1065

\bibitem[\protect\citeauthoryear{{Negueruela} \& {Marco}}{{Negueruela} \&
  {Marco}}{2012}]{negueruela2012}
{Negueruela} I.,  {Marco} A.,  2012, \mn@doi [\aj]
  {10.1088/0004-6256/143/2/46}, 143, 46

\bibitem[\protect\citeauthoryear{{Negueruela}, {Gonz\'{a}lez-Fern\'{a}ndez},
  {Marco}  \& {Clark}}{{Negueruela} et~al.}{2011}]{negueruela2011}
{Negueruela} I.,  {Gonz\'{a}lez-Fern\'{a}ndez} C.,  {Marco} A.,   {Clark}
  J.~S.,  2011, \mn@doi [\aap] {10.1051/0004-6361/201016102}, 528, A59

\bibitem[\protect\citeauthoryear{{Negueruela}, {Mongui{\'o}}, {Marco},
  {Tabernero}, {Gonz{\'a}lez-Fern{\'a}ndez}  \& {Dorda}}{{Negueruela}
  et~al.}{2018}]{negueruela2018}
{Negueruela} I.,  {Mongui{\'o}} M.,  {Marco} A.,  {Tabernero} H.~M.,
  {Gonz{\'a}lez-Fern{\'a}ndez} C.,   {Dorda} R.,  2018, \mn@doi [\mnras]
  {10.1093/mnras/sty718}

\bibitem[\protect\citeauthoryear{{Reid}, {Dame}, {Menten}  \&
  {Brunthaler}}{{Reid} et~al.}{2016}]{reid2016}
{Reid} M.~J.,  {Dame} T.~M.,  {Menten} K.~M.,   {Brunthaler} A.,  2016, \mn@doi
  [\apj] {10.3847/0004-637X/823/2/77}, 823, 77

\bibitem[\protect\citeauthoryear{{Riess} et~al.,}{{Riess}
  et~al.}{2018}]{riess2018}
{Riess} A.~G.,  et~al., 2018, preprint (\mn@eprint {arXiv} {1801.01120})

\bibitem[\protect\citeauthoryear{{Rivinius}, {Carciofi}  \&
  {Martayan}}{{Rivinius} et~al.}{2013}]{rivinius2013}
{Rivinius} T.,  {Carciofi} A.~C.,   {Martayan} C.,  2013, \mn@doi [\aapr]
  {10.1007/s00159-013-0069-0}, 21, 69

\bibitem[\protect\citeauthoryear{{Sagar} \& {Griffiths}}{{Sagar} \&
  {Griffiths}}{1998}]{sagar1998}
{Sagar} R.,  {Griffiths} W.~K.,  1998, \mn@doi [\mnras]
  {10.1046/j.1365-8711.1998.01551.x}, 299, 1

\bibitem[\protect\citeauthoryear{{Scargle}}{{Scargle}}{1982}]{scargle1982}
{Scargle} J.~D.,  1982, \mn@doi [\apj] {10.1086/160554}, 263, 835

\bibitem[\protect\citeauthoryear{{Soszy\'{n}ski} et~al.,}{{Soszy\'{n}ski}
  et~al.}{2008}]{soszynski2008}
{Soszy\'{n}ski} I.,  et~al., 2008, \actaa, 58, 163

\bibitem[\protect\citeauthoryear{{Stassun} \& {Torres}}{{Stassun} \&
  {Torres}}{2018}]{stassun2018}
{Stassun} K.~G.,  {Torres} G.,  2018, preprint (\mn@eprint {arXiv}
  {1805.03526})

\bibitem[\protect\citeauthoryear{{Stellingwerf}}{{Stellingwerf}}{1978}]{stelli%
ngwerf1978}
{Stellingwerf} R.~F.,  1978, \mn@doi [\apj] {10.1086/156444}, 224, 953

\bibitem[\protect\citeauthoryear{{Subramaniam}, {Carraro}  \&
  {Janes}}{{Subramaniam} et~al.}{2010}]{subramaniam2010}
{Subramaniam} A.,  {Carraro} G.,   {Janes} K.~A.,  2010, \mn@doi [\mnras]
  {10.1111/j.1365-2966.2010.16345.x}, 404, 1385

\bibitem[\protect\citeauthoryear{{Tabernero}, {Dorda}, {Negueruela}  \&
  {Gonz{\'a}lez-Fern{\'a}ndez}}{{Tabernero} et~al.}{2018}]{tabernero2018}
{Tabernero} H.~M.,  {Dorda} R.,  {Negueruela} I.,
  {Gonz{\'a}lez-Fern{\'a}ndez} C.,  2018, \mn@doi [\mnras]
  {10.1093/mnras/sty399}, 476, 3106

\bibitem[\protect\citeauthoryear{{Tadross}}{{Tadross}}{2008}]{tadross2008}
{Tadross} A.~L.,  2008, \mn@doi [\mnras] {10.1111/j.1365-2966.2008.13554.x},
  389, 285

\bibitem[\protect\citeauthoryear{{Xu} et~al.,}{{Xu} et~al.}{2013}]{xu2013}
{Xu} Y.,  et~al., 2013, \mn@doi [\apj] {10.1088/0004-637X/769/1/15}, 769, 15

\bibitem[\protect\citeauthoryear{{Zhang}, {Reid}, {Menten}, {Zheng},
  {Brunthaler}, {Dame}  \& {Xu}}{{Zhang} et~al.}{2013}]{zhang2013}
{Zhang} B.,  {Reid} M.~J.,  {Menten} K.~M.,  {Zheng} X.~W.,  {Brunthaler} A.,
  {Dame} T.~M.,   {Xu} Y.,  2013, \mn@doi [\apj] {10.1088/0004-637X/775/1/79},
  775, 79

\makeatother
\end{thebibliography}

\bsp
\label{lastpage}
\end{document}